\documentclass[preprints,article,accept,moreauthors]{Definitions/mdpi} 

\firstpage{1} 
\pubvolume{28}
\issuenum{11}
\articlenumber{1258}
\pubyear{2026}
\copyrightyear{2026}
\externaleditor{Jian-Wei Pan, Artur Ekert, Barry C. Sanders and  Feihu Xu} 
\datereceived{19 October 2026} 
\daterevised{10 November 2026} 
\dateaccepted{12 November 2026} 
\datepublished{14 November 2026} 
\pdfoutput=1 

\usepackage{physics}
\usepackage{siunitx}
\usepackage{overpic}

\usetikzlibrary{arrows.meta,positioning,fit,calc}
\setlist[itemize]{leftmargin=*}
\setlist[enumerate]{leftmargin=*}
\Title{From BBM92 Measurements to Secure Keys: Challenges in a Practical Implementation} 

\Author{Karolina 
 P. Schatz~$^{1,2,}$*\orcidA{}, 
Elia Falcioni~$^{1,3,4}$\orcidB{}, 
Panagiotis Papanastasiou~$^{1,2,5}$\orcidC{}
and Marco Lucamarini~$^{1,2,}$*\orcidD{}}

\AuthorNames{Karolina P. Schatz, Elia Falcioni, Panagiotis Papanastasiou and Marco Lucamarini}

\address{%
$^{1}$ \quad School of Physics, Engineering and Technology, University of York, York YO10 5FT, UK
\\
$^{2}$ \quad York Centre for Quantum Technologies, University of York, York YO10 5FT, UK\\
$^{3}$ \quad School of Science and Technology, University of Camerino, I-62032 Camerino, Italy\\
$^{4}$ \quad School of Advanced Studies ``Carlo Urbani'', University of Camerino, I-62032 Camerino, Italy\\
$^{5}$ \quad Department of Informatics and Telecommunications, National and Kapodistrian University of Athens, \linebreak  157 84~Athens
, Greece}

\corres{Correspondence: karolina.schatz@york.ac.uk (K.P.S.); marco.lucamarini@york.ac.uk (M.L.)}

\abstract{The protocol introduced by C. H. Bennett, G. Brassard and N. D. Mermin in 1992 
 (BBM92) is one of the most important and widely adopted quantum key distribution (QKD) protocols. As with any other QKD protocol, its practical security depends on satisfying the assumptions underlying its theoretical security proofs. Here, we examine two common implementations of BBM92: a conventional passive receiver and a time-multiplexed configuration that halves the number of detectors. We characterise the passive polarisation analysers common to both architectures and present experimental results obtained with the time-multiplexed receiver, employing a continuous-wave (CW) pumped spontaneous parametric down-conversion (SPDC) source of polarisation-entangled photon pairs and threshold detectors. We investigate the challenges encountered during post-processing when attempting to satisfy the requirements of existing security proofs and discuss the gap between theoretical security assumptions and the extraction of a provably secure cryptographic key.}


\begin{document}

\section{Introduction}

Quantum key distribution (QKD) emerged from the pioneering work of Charles H. Bennett and Gilles Brassard, who transformed Stephen Wiesner's earlier ideas on conjugate coding into a cryptographic protocol for establishing a secret key over an insecure communication channel. Introduced in 1984, their protocol, now universally known as BB84~\citep{bennett2014}, demonstrated how the laws of quantum mechanics could be exploited to distribute secret keys without relying on assumptions about the computational capabilities of a potential eavesdropper. A~fascinating historical account of these developments can be found in Brassard's personal recollections~\citep{briefhistory}.

Independently, in~1991, Artur Ekert proposed a quantum cryptographic protocol based on entangled particles~\citep{E91}, introducing a connection between QKD and the violation of Bell inequalities. The~following year, Bennett, Brassard and Mermin introduced BBM92~\citep{bennett1992}, demonstrating that entanglement-based QKD could also be realised without testing a Bell inequality, relying instead on measurements in two complementary bases. An~important feature of the entanglement-based approach is that the entanglement source need not be trusted and may even be controlled by an eavesdropper, provided that the legitimate parties' measurement devices satisfy the assumptions of the security analysis. This makes BBM92 particularly attractive for quantum communication networks, where a central source may distribute entanglement to distant~users.

Since its introduction, BBM92 has become one of the most established entanglement-based QKD protocols, owing to its conceptual simplicity and its compatibility with practical photonic implementations. Early demonstrations using entangled photons~\citep{jennewein2000,tittel2000} paved the way for increasingly sophisticated systems, including multi-user metropolitan quantum networks~\citep{joshi2020} and long-distance entanglement distribution over optical fibres. Notable achievements include the distribution of polarisation entanglement over 248~km of deployed telecommunication fibre~\citep{neumann2022} and entanglement-based QKD over 404~km of optical fibre~\citep{zhuang2025}. In~our earlier work, we investigated the feasibility of quantum communications over a 224~km submarine pair of optical fibres connecting the United Kingdom and Ireland, characterising its phase and polarisation stability and demonstrating photon-pair coincidences after propagation through a total distance of 448~km in fibre channels~\citep{amiesking2023}. These developments illustrate both the potential and the practical challenges of entanglement-based quantum communications over existing optical infrastructure, particularly for networks in which entanglement is distributed to distant users without requiring a trusted source or intermediate key-relay~nodes.

Despite this experimental progress, guaranteeing the security of a practical implementation remains more challenging than demonstrating its basic~operation. 

Rigorous security proofs have developed over decades~\citep{lochau1999,shor2000,KoashiPreskill2003,GLLP2004,Tomamichel2012,TomamichelLeverrier2017}, including treatments of finite-size effects and non-ideal photodetection. Nevertheless, important challenges remain in connecting these theoretical guarantees to practical implementations~\citep{zapatero2025,Tupkary2026Review}, and security arguments can contain gaps arising from discrepancies between protocol specifications, proof assumptions and the treatment of experimental imperfections.
The limitations introduced by realistic devices were recognised early on: in their seminal work on practical quantum cryptography, Brassard and collaborators showed that transmission loss, imperfect photon sources and realistic detectors could undermine security claims based on idealised models~\citep{brassard2000,Fung2009Mismatch}. Subsequent studies demonstrated attacks exploiting detector-response mismatches and nonlinearities, including time-shift attacks and detector blinding~\citep{qi2007,lydersen2010}. Detection-time-dependent information leakage in BBM92-like protocols has likewise been demonstrated~\citep{LamasLinares2007} and generalised to an implementation security problem~\citep{scarani2014black}. Together, these investigations highlight an essential distinction: the security of an abstract QKD protocol does not automatically extend to its physical implementation unless the assumptions of the proof accurately describe the apparatus and the post-processing~\mbox{procedure}.

This issue is particularly relevant to practical BBM92 systems. Passive basis selection commonly relies on beam splitters whose actual splitting ratios may differ from their nominal values; unequal optical transmission and detector efficiencies may introduce basis- and outcome-dependent detection probabilities. Threshold detectors cannot resolve photon number, and~double clicks or more complex multi-click patterns require an explicit treatment~\citep{Beaudry2008Squashing, Koashi2008Practical, TsurumaruTamaki2008, FungChauLo2011, Gittsovich2014}. Recent developments have addressed imperfect threshold detectors~\citep{nahar2026}, passive detection with imperfections and memory effects~\citep{Wang2025Passive}, and~passive entanglement-based QKD with biased basis choice, including BBM92~\citep{kawakami2026}. Nevertheless, the~applicability of a particular proof depends on the physical measurement model and the precise classical processing used. In~particular, identifying coincidences and publicly communicating detection-time information can raise security questions that do not appear in idealised protocol~descriptions.

In this work, we revisit an apparently straightforward implementation of the BBM92 protocol, using established laboratory techniques. We employ a CW-pumped spontaneous parametric down-conversion source of polarisation-entangled photons, passive analysers and superconducting nanowire single-photon detectors (SNSPDs). We examine the practical imperfections and processing decisions affecting the applicability of existing security proofs, including biased basis selection, detection-efficiency mismatch, multiple clicks and the identification of measurement outcomes from coincidence timing. We do not intend to propose a new security proof but to illustrate the gaps that have to be simultaneously included in the theory to guarantee a secure implementation and operation of QKD based on the BBM92 protocol.
Our contribution therefore is an experimentally grounded assessment of the correspondence between the physical BBM92 receiver, its classical processing and the assumptions of existing security proofs. Unlike earlier demonstrations of CW passive-basis BBM92 with two detectors per user~\citep{joshi2020}, and~detector-saving time multiplexing in entanglement-based QKD~\citep{Kaltwasser2024}, we focus on the security implications of our specific implementation: the basis is identified by the physical detector, whereas the bit pair is reconstructed from joint timing peaks. Rather than proposing a new protocol or claiming that previous demonstrations are insecure, we identify the additional conditions that would be needed to certify a key from our measured~data.

\section{Materials 
 and~Methods}
\subsection{BBM92 Reference~Setup}
\begin{figure}[H]
\centering
    \begin{overpic}[width=0.85\linewidth]{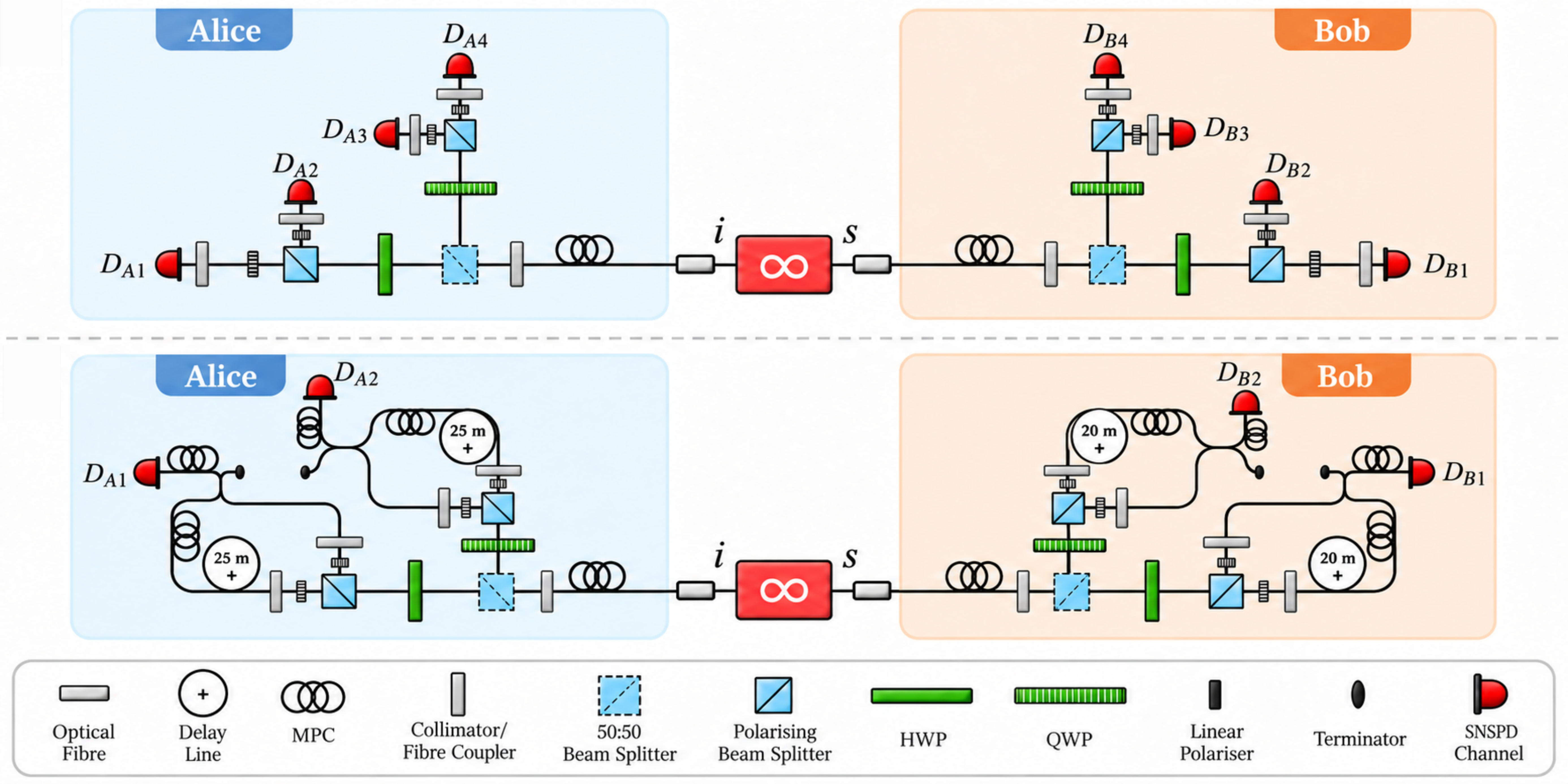}
        \put(0,47){(\textbf{a})}
        \put(0,25){(\textbf{b})}
    \end{overpic}
    \vspace{6pt}
\caption{Schematic 
 of the two passive BBM92 receiver architectures considered in this work. The~central infinity symbol represents an entangled-photon source (EPS) emitting entangled photons in modes $i$ (idler), launched towards Alice, and~$s$ (signal), launched towards Bob. (\textbf{a}) Traditional BBM92 setup using a dedicated detector channel for each possible outcome. (\textbf{b}) Configuration used in this work. Both bit outputs of each basis are multiplexed onto one detector channel with a characteristic time delay between different bit values. Bit values can only be identified in coincidence based on their relative arrival~{times}.}
\label{fig:setup}
\end{figure}

BBM92 is well known, and~we refer to the original paper~\citep{bennett1992} for a detailed explanation. Figure~\ref{fig:setup}a illustrates a reference implementation of passive BBM92, employing four detectors per user, one for each possible measurement outcome. Although~our experiment adopts the two-detector configuration shown in Figure~\ref{fig:setup}b, both configurations share the same entangled-photon source and passive polarisation analysers. We therefore begin by examining the practical imperfections of these common optical components before~considering the additional challenges introduced by time~multiplexing.

    \vspace{6pt}

Ideally, BBM92 utilises the following entangled-photon pair emitted by the central SPDC source: 
\begin{equation}
\ket{\Phi^+}_{AB}
=
\frac{
\ket{H}_A\ket{H}_B+
\ket{V}_A\ket{V}_B
}{\sqrt{2}} ,
\label{eq:phi-plus}
\end{equation}
where $\ket{H},\ket{V}$ are eigenvectors of the basis $Z$, associated with horizontal and vertical polarisation, respectively. The~eigenvectors of the $X$ basis are obtained as $\ket{\pm}=(\ket{H}\pm\ket{V})/\sqrt{2}$ and are associated with diagonal and anti-diagonal polarisation. 
The subscripts in Equation~\eqref{eq:phi-plus} refer to the separate users and measurement stations `Alice' and `Bob,' and we denote users, bases and resulting bit values as
\begin{equation}
U\in\{A,B\},\qquad
W\in\{Z,X\},\qquad
b\in\{0,1\}. 
\end{equation}

Each user ideally receives one photon of the entangled pair and measures its polarisation in a randomly chosen basis $W$, and the user~subsequently maps the result to a bit value $b$. 
At each user $U$, numbered threshold detectors $D_{Un}$, with~$n\in\{1,2,3,4\}$, are used to record single-photon events, enabling the identification of coincident single-photon detections at $A$ and $B$ as well as multi-detection events within a single~user. 

Via a public, classical communication channel, Alice and Bob perform basis reconciliation to obtain a sifted key. They then use a pre-agreed procedure for parameter estimation, error correction, and~privacy amplification to arrive at a secret~key.

\subsubsection{Continuous-Wave Entangled-Photon~Source}
In our experiment, we adopt an entangled-photon source (EPS) manufactured by OZ Optics~\citep{OZOpticsRuby} (Ottawa, Canada), to~whose customisation we contributed. The~source employs a CW-pump laser that drives two periodically poled nonlinear crystals arranged in an auto-balanced Mach–Zehnder interferometer configuration~\citep{HornJennewein2019}. The~resulting polarisation-entangled signal and idler photons are coupled into separate optical~fibres. 

This EPS presents multiphoton emissions (see Section~\ref{sec:multiplex} for more details) and a broad optical spectrum with a full width at half maximum of approximately \SI{60}{\nano\metre}, centred in the telecom C-band at 1560~nm. To~reduce chromatic dispersion for longer transmission channels and to circumvent a manufacturing defect in the source that increases the background noise near the centre of the spectrum, we select signal photons at 1550.1~nm and idler photons at 1569.3~nm via optical filters with pass-bands of $\Delta_f$ = 1.391~nm each. This yields a good extinction ratio between photon pairs and background counts at the price of source brightness, resulting in a singles detection rate on the order of 1~MHz per output channel, a~detected coincidence rate of $45.7 \times 10^3$~s$^{-1}$, and~a heralding efficiency of approximately 5\% per arm. The~experiment was performed on an optical table, with~a few metres of optical fibre between the source and the receivers and no long-distance transmission~link.

\subsubsection{Polarisation Analysers and Passive Basis~Selection}
The polarisation analysers (PAs), common to both configurations in Figure~\ref{fig:setup}, are in-house-built, fixed-assembly research devices designed for flexible polarisation measurements while maintaining long-term coupling stability.~They perform the relevant polarisation operations in free space, using a nominally 50:50 beam splitter (BS) for passive basis selection, followed by waveplates and polarising beam splitters (PBSs) for polarisation analysis.
An additional waveplate is present in the $Z$-basis path that is technically not needed for BBM92, but~it in practice contributes to equalise the optical paths. Linear polarisers (LPs) are added before each fibre coupler to clean PBS's outputs, especially the reflected mode, which typically includes a non-negligible orthogonal component and can lead to a increased quantum bit error rate (QBER) if not further isolated. While additional LPs introduce uneven losses to each photon path, the~improved polarisation state isolation and reliably low QBER outweigh this~downside.

The BSs used in this setup are commercial-grade components and exhibit some variability in their splitting ratios, as~is commonly encountered in off-the-shelf optics. We measure that the BS located at $B$ results in an even splitting ratio of an incoming photon beam, while the one at $A$ measures a splitting ratio closer to 47:53 reflected vs. transmitted.
Importantly for the security analysis, the~applicability of a proof depends on the assumed basis-selection mechanism and the complete optical measurement. A~perfectly balanced 50:50 splitter is required by some passive squashing constructions~\citep{Gittsovich2014}, but~biased passive basis selection can also be treated under suitable assumptions~\citep{kawakami2026}.
The concern is therefore not simply a deviation from equal basis probabilities, but~whether the basis selection and detection
efficiencies depend on optical properties that may be influenced by an eavesdropper (Eve). In~particular, wavelength-dependent beam-splitter responses have been exploited experimentally in passive-basis polarisation-encoded QKD~\citep{Li2011Wavelength}.
Although that attack was not explicitly considered during the security analysis of our BBM92 receiver, it illustrates the importance of characterising the
accepted spectral range and the out-of-band response. Optical filtering can constrain these effects, but~an ideal filter response should not be assumed without~justification.

\begin{figure}[H]
\centering
    \begin{minipage}{0.4\linewidth}
        \begin{overpic}[width=\linewidth]{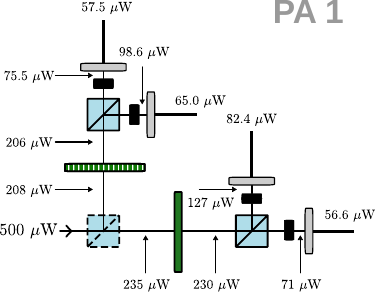}
            \put(0,74){(\textbf{a})}
        \end{overpic}
    \end{minipage}
    \hspace{1em}
    \begin{minipage}{0.4\linewidth}
        \begin{overpic}[width=\linewidth]{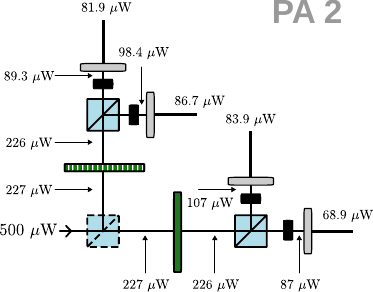}
            \put(0,74){(\textbf{b})}
        \end{overpic}
    \end{minipage}
    \includegraphics[width=0.45\linewidth]{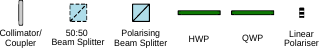}
    \vspace{6pt}
    \caption{(\textbf{a}) PA1, Alice; (\textbf{b}) PA2, Bob. Optical 
    transmission characterisation of the passive polarisation analysers employed in both configurations of Figure~\ref{fig:setup}. The~measurements illustrate path-dependent losses that are present independently of the time-multiplexing~stage.}
    \label{fig:PAs_losses}
\end{figure}

Figure~\ref{fig:PAs_losses} presents an experimental characterisation of the optical transmission through the passive polarisation analysers illustrated in Figure~\ref{fig:setup}a. The~measurements are performed using a calibrated power metre and injecting a diagonally polarised laser beam into the PAs, with~all the waveplates set to zero degrees, to~provide a common calibration configuration.
The four measured output powers sum to 261.5~\mbox{\textmu W} 
 for PA~1 and 321.4~\mbox{\textmu W} for PA~2, compared with the 500~\mbox{\textmu W} input power at each analyser. These correspond to optical transmissions of 52.3\% and 64.3\%, or~insertion losses of approximately 2.81~dB and 1.92~dB, respectively. These calibration values exclude subsequent multiplexing components and the SNSPD detection efficiencies.
Significant differences in transmission are observed between the various optical paths, even before accounting for the additional losses introduced by fibre connections, detector efficiencies and, in~our next implementation, time multiplexing. These asymmetries are intrinsic to passive analysis units of the conventional four-detector-per-user configuration. 
While experimentalists may attempt to compensate for these imperfections by equalising the transmission efficiencies of the different measurement paths, this is not an easy task in~practice.

\vspace{10pt}

Importantly, security proofs may differ in their assumptions about the receiver from the experimentally achievable devices. 
For example, conventional squashing models for passive BB84 measurements consider balanced basis selection with a 50:50 beam splitter~\cite{Gittsovich2014}, while the prepare-and-measure analysis of Tomamichel and Leverrier requires the probability of an inconclusive outcome to be independent of the measurement basis for any given input state~\cite{TomamichelLeverrier2017}. Other results relax restrictions on passive basis selection~\cite{Wang2025Passive,kawakami2026}.

One possible approach to balancing the observed count rates is to randomly discard a fraction of the detection events in the more efficient paths. Alternatively, individual fibre couplers can be deliberately misaligned to introduce additional optical losses and achieve more uniform detection rates. Both approaches effectively reduce the usable data towards the level of the least efficient path, potentially decreasing the secret key rate significantly. Moreover, equalising the observed count rates through classical post-processing does not necessarily reproduce the physical measurement model required by a security proof, particularly in the presence of multiphoton events. Such countermeasures may therefore be unattractive to commercial QKD providers, who are understandably reluctant to sacrifice system performance to satisfy additional security~assumptions.

A further difficulty arises from imperfect alignment of the polarisation measurement bases. Even if the optical losses are equalised, Alice's and Bob's analysers may not realise exactly the intended complementary $Z$ and $X$ bases. Ideally, their eigenstates satisfy $|\langle z|x\rangle|=1/\sqrt{2}$ for any pair of eigenstates $|z\rangle$ and $|x\rangle$ belonging to different bases. In~terms of linear polarisation, this corresponds to a relative angle of $45^\circ$ between the two measurement bases. Measurements should therefore be performed at the beginning of the communication session, and~potentially monitored during operation, to~verify that the polarisation axes remain correctly aligned. Such characterisation may require additional equipment and calibration procedures, increasing the cost and complexity of the~system.

More generally, even with careful engineering and calibration, achieving ideal specifications such as a perfectly balanced 50:50 beam splitter or perfectly aligned measurement bases is extremely challenging experimentally. Small deviations from these ideal conditions are practically unavoidable and must therefore be explicitly considered when establishing the security of a realistic~implementation.

These observations illustrate that even a conventional passive BBM92 receiver with four detectors per user departs from the idealised measurement model commonly adopted in security analyses. In~particular, passive basis imbalance, unequal transmission and detection efficiencies, imperfect basis alignment, and multiple detection events must all be considered when selecting an applicable security proof. The~practical implementation of BBM92 therefore already presents nontrivial security challenges before any further reduction in detector resources is attempted. The~relevance of selected security results to both the conventional and time-multiplexed implementations is summarised in Table~\ref{tab:security_proofs_combined}.

\begin{table}[H]
\small
\setlength{\tabcolsep}{3pt}
\caption{Selected security proofs for the conventional passive receiver (Figure~\ref{fig:setup}a) and the time-multiplexed receiver (Figure~\ref{fig:setup}b). No result alone establishes composable finite-key security for the complete apparatus and~processing.}
\label{tab:security_proofs_combined}

\begin{tabularx}{\textwidth}{>{\raggedright\arraybackslash}p{3cm}LLL}
\toprule
\textbf{Reference} & \textbf{Main Scope or Condition} & \textbf{Conventional Receiver (a)} & \textbf{Time-Multiplexed Receiver~(b)} \\
\midrule
Koashi~et~al.~\citep{Koashi2008Practical} &
Asymptotic BBM92; multiphoton inputs, threshold detection and active basis choice. &
Passive basis choice and efficiency mismatch {\linebreak} need justification. &
No delayed-bit reconstruction or detector~recovery. \\
\midrule
Tsurumaru--Tamaki~\citep{TsurumaruTamaki2008}~~~~~~~~~ &
Squashing for active-basis BBM92; random assignment of double clicks. &
Passive cross-basis clicks not~covered. &
Shared-detector outcomes and timing not~covered. \\

\midrule
Gittsovich~et~al.~\citep{Gittsovich2014} &
Optical squashing for balanced passive BB84; multiphotons and cross-clicks. &
Biased splitter and path-dependent losses fall outside the specified model. &
Delay encoding and recovery not~modelled. \\
\midrule
Joshi~et~al.~\citep{joshi2020} &
Two-detector BBM92; 3.7~ns basis delay, finite-key analysis and timing-tagging bound; secret-key rate assumes untagged events. &
Different detection and basis-reconciliation procedure. &
Closest precedent; short-delay recovery and suppressed multiclicks not explicitly~treated. \\
\midrule
Tomamichel--Leverrier~\citep{TomamichelLeverrier2017} &
Composable finite-key proof; arbitrary finite-dim states and basis overlap. Binary outcomes for entanglement-based model. 
&
Loss postselection and optical-to-effective measurement reduction require justification. &
Also requires timing-announcement and detector-memory~treatment. \\
\midrule
Lim~et~al.~\citep{Lim2021} &
Composable finite-key BBM92 with specified effective measurements. &
Optical reduction and click rules not established. &
Joint timing-derived bits not~covered. \\
\midrule
Kawakami~et~al.~\citep{kawakami2026} &
Asymptotic passive BBM92; biased basis choice, cross-click bounds and specified detector efficiencies. &
Within-basis efficiency assumptions.
& Timing reconstruction and recovery not~treated. \\

\bottomrule
\end{tabularx} 

\end{table}

\begin{table}[H]\ContinuedFloat
\small
\caption{\textit{Cont.}}
\setlength{\tabcolsep}{3pt}
\begin{tabularx}{\textwidth}{>{\raggedright\arraybackslash}p{3cm}LLL}
\toprule
\textbf{Reference} & \textbf{Main Scope or Condition} & \textbf{Conventional Receiver (a)} & \textbf{Time-Multiplexed Receiver~(b)} \\

\midrule
Wang~et~al.~\citep{Wang2025Passive} &
Finite-size phase-error estimation for passive detection; characterised imperfections and specified detector memory~effects. &
Promising framework for incorporating realistic BBM92 in finite-size security analysis. &
Recovery-induced memory; potential application to timing-based outcome~reconstruction. \\
\midrule
Nahar~et~al.~\citep{nahar2026} &
Reduction framework for imperfect threshold detectors, with~detection efficiencies and dark counts. &
Provides tools for connecting characterised optical setup to measurements in security analyses. &
Foundation for shared physical detectors; potential extension to timing-dependent outcome~reconstruction. \\
\bottomrule
\end{tabularx}
\end{table}


\subsection{Two-Detector Time-Multiplexed~Receiver}
\label{sec:multiplex}
The experimental configuration employed to acquire data in this work is shown in Figure~\ref{fig:setup}b. It retains the source and passive polarisation analysers discussed above but~introduces time multiplexing to reduce the number of physical detectors from four to two per user. Since single-photon detectors, particularly SNSPDs, represent a substantial fraction of the cost of a QKD implementation~\citep{FDL+13}, time multiplexing is an established approach to reducing hardware requirements~\citep{joshi2020,Kaltwasser2024}.

In our implementation, the~two polarisation outputs corresponding to bit values 0 and 1 in each measurement basis are combined onto a single SNSPD channel, using fibre delay lines to distinguish the outcomes by their arrival times. This reduces the number of physical detection channels from eight to four and promotes the similarity between the two bit value detections within each basis. Before~acquisition, each incident polarisation mode was manually adjusted at the SNSPD to maximise its detection efficiency. The~resulting characterised efficiencies per SNSPD are listed in Table~\ref{tab:receiver_parameters}.~Nevertheless, the~complete efficiencies of the two logical outcomes can differ because they traverse distinct optical paths, including the multiplexing delays. Using one physical SNSPD for both outcomes reduces the number of separate detector-response functions, but~it does not establish equality of the effective logical-outcome efficiencies or a general bound on Eve's ability to exploit them. The~reduction in detector resources therefore comes with additional optical losses and more complex reconstruction of detection~events.

\begin{table}[H]

\caption{Principal parameters of the laboratory receiver. Optical-analyser transmissions are calculated from Figure~\ref{fig:PAs_losses}. SNSPD efficiencies were measured following manual optimisation of the incident polarisation and are not the end-to-end efficiencies of the logical measurement~paths.}
\label{tab:receiver_parameters}

\begin{tabularx}{\textwidth}{lL}
\toprule
\textbf{Parameter} & \textbf{Value} \\
\midrule
Source-to-receiver fibres & A few metres on an optical table; no long-distance~link \\
Basis-selection beam splitters & Alice: approximately 47:53 (reflected:transmitted); \newline Bob: approximately 50:50 \\
PA~1 optical transmission & 52.3\% (2.81~dB insertion loss) \\
PA~2 optical transmission & 64.3\% (1.92~dB insertion loss) \\
SNSPD efficiencies, channels 3 and 4 & $(87\pm3)\%$ and $(90\pm1)\%$, respectively \\
SNSPD efficiencies, channels 5 and 6 & $(87\pm1)\%$ and $(93\pm1)\%$, respectively \\
Alice's multiplexing delays, $X$ and $Z$ & $(108,807.9\pm1)$ 
~ps and $(108,872.6\pm3)$~ps \\
Bob's multiplexing delays, $X$ and $Z$ & $(133,388.2\pm1)$~ps and $(133,304.6\pm1)$~ps \\
Coincidence window & 150~ps \\
\bottomrule
\end{tabularx}
\end{table}

In implementing the multiplexing of two or more detection outputs into a single detector, care is required to not overlook the impact of detector recovery time in such setups. Immediately after a detection event, threshold detectors require a certain period of time until the photosensitive state is re-established and another detection can be made. Based on our characterisation of same-detector consecutive click-pair rates, our SNSPDs take between \SI{65}{\nano\second} and \SI{89}{\nano\second} to recover \SI{50}{\percent} of their maximum efficiency. We therefore introduce delays of at least 100~ns between the two logical bit outcomes within each basis. These delays allow substantial detector recovery and increase the probability of resolving temporally separated detections arising from multiphoton events or dark~counts.

\begin{figure}[H]
    \begin{minipage}[t]{0.6\textwidth}
    \vspace{0pt}
    \centering
        \begin{overpic}[width=0.95\linewidth]{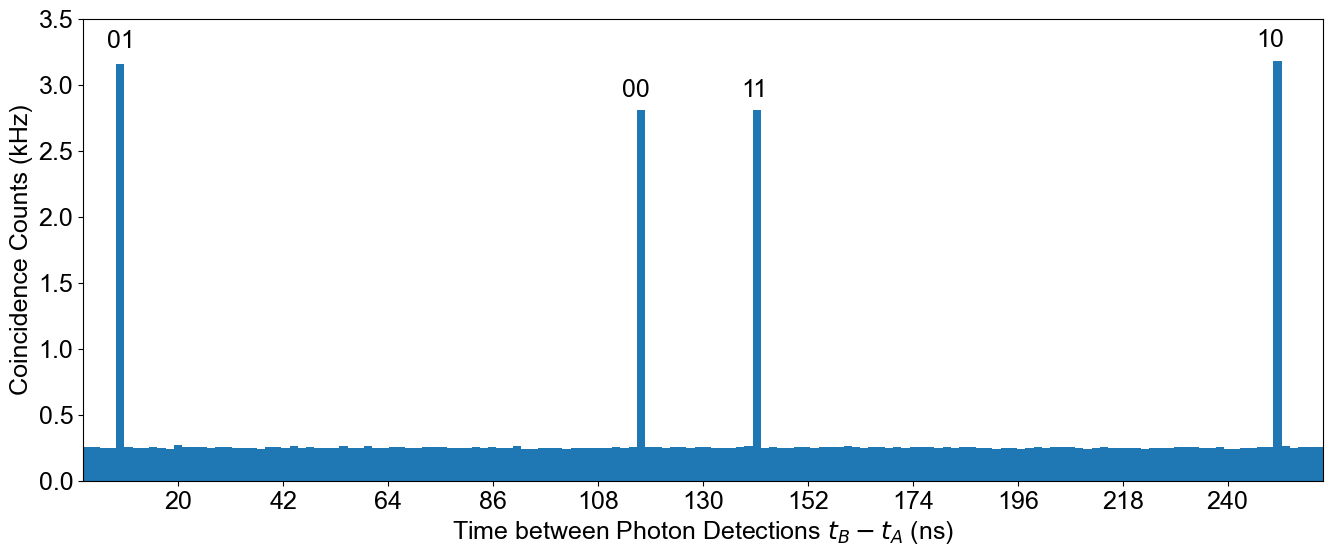}
             \put(-5,38){(\textbf{a})}
        \end{overpic}
    \vspace{0.5cm}
        \begin{overpic}[width=0.95\linewidth]{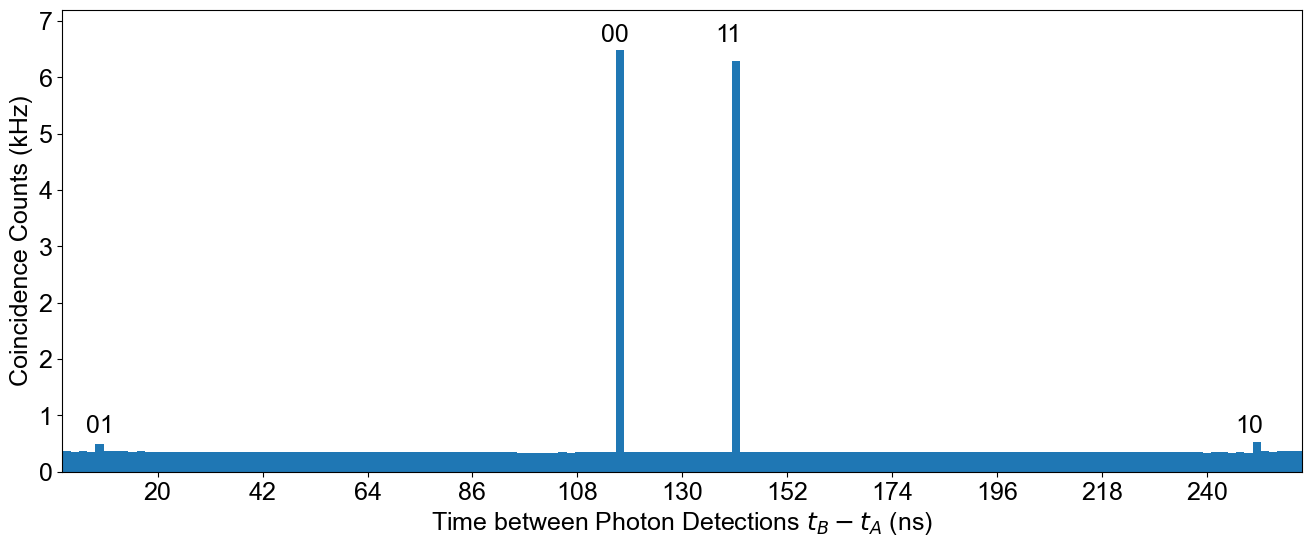}
             \put(-5,38){(\textbf{b})}
        \end{overpic}
    \end{minipage}%
    \hfill
    \begin{minipage}[t]{0.4\textwidth}
        \vspace{0pt}
        \centering
        \begin{overpic}[width=0.95\linewidth]{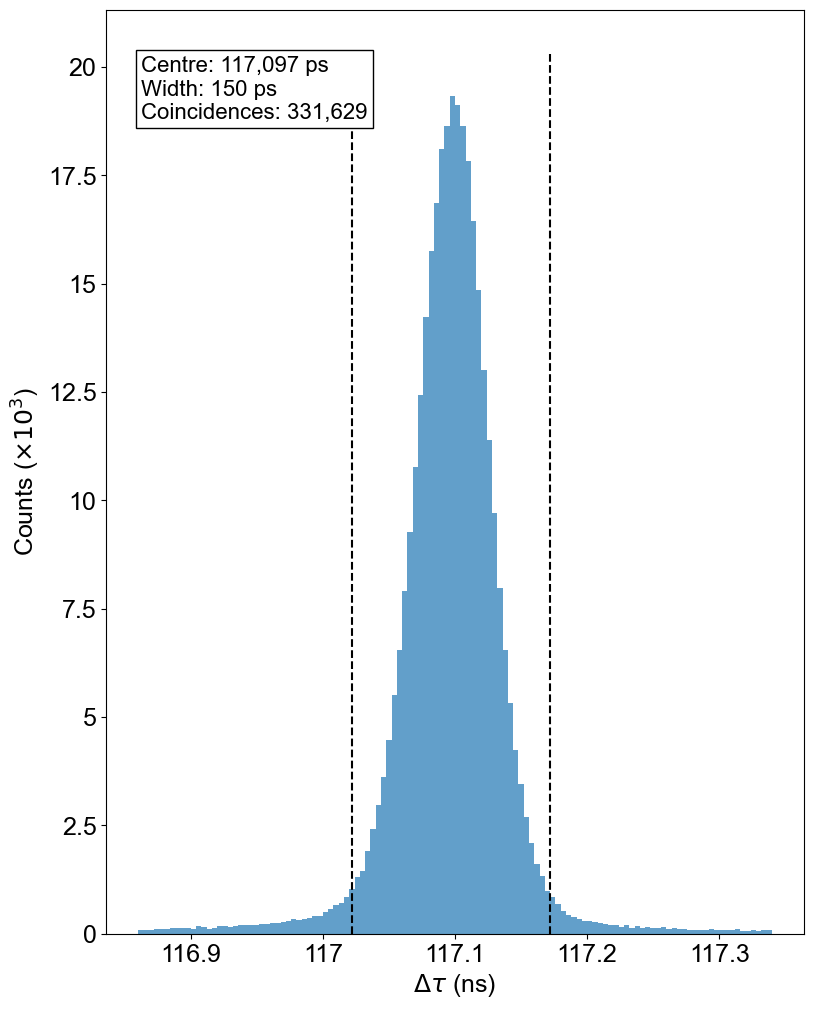}
             \put(-3,96){(\textbf{c})}
        \end{overpic}
    \end{minipage}
    \caption{(\textbf{a}) 
 Coincidence distribution measured in mismatched bases ($A_Z$, $B_X$). All four peaks are clearly visible. (\textbf{b}) Coincidence distribution measured in matching bases ($A_X$, $B_X$). Mainly correlated polarisation detections occur. (\textbf{c}) Individual coincidence peak for 00 detections in $A_Z$, $B_Z$, showing the coincidence window of 150~ps, determined by detector timing jitter. At~longer transmission channel lengths, chromatic dispersion would necessitate a wider coincidence~window.}
    \label{fig:coincidence_peaks}
\end{figure}

A related two-detector architecture was employed in~\citep{joshi2020}, which features a basis-dependent optical delay of approximately 3.7~ns.~Although~this delay is sufficient to distinguish the coincidence peaks, it is much shorter than typical SNSPD recovery times~\citep{PhotonSpotDetectors}. Consequently, two photons arriving at the same detector through different optical paths may not produce two distinguishable detection events. This can suppress multiple-click patterns that are relevant to security analyses of multiphoton inputs. While Ref.~\citep{joshi2020} provides a finite-key security analysis accounting for passive basis selection and certain timing effects, the~impact of detector recovery on multiple-click statistics is not explicitly addressed. In~our implementation, the~delays exceed 100~ns, substantially reducing this problem and allowing a more complete reconstruction of multiple-click events.
Figure~\ref{fig:coincidence_peaks} depicts example distributions in time of coincident polarisation events measured between Alice and Bob in our setup. Each of the four two-photon polarisation modes has a distinct peak location, with~a minimum of 100~ns between the expected possible detections on $A$ involving the same photon on $B$ and vice~versa.

In principle, an~appropriate timing reference associated with photon-pair generation could allow each user to distinguish the delayed and undelayed logical outcomes locally. For~example, a~pulsed SPDC source could provide a pump-derived timing reference. The~present source is CW-pumped and provides no such event reference. Consequently, the~logical outcomes are reconstructed from the complete joint timing dataset rather than being independently available to Alice and Bob. This arrangement is suitable for characterising the expected BBM92 correlations but~cannot be assumed to implement a secure key-distribution protocol without an additional analysis of the required classical communication. Likewise, the~measured coincidence statistics cannot be assumed to remain unchanged under modifications to the source or detector architecture without further experimental~verification.

\section{Results}
\subsection{Experimental Results and Multiphoton~Statistics}

Even if not cryptographically suitable, the~present implementation represents a direct folded version of the BBM92 protocol and reproduces the same statistics. Hence, we use it to characterise its performance. Introducing a suitable timing reference, such as that provided by a pulsed source, could address the local identification of logical outcomes, but~it would not by itself establish security without a compatible analysis of the receiver and its post-processing. From~the differences $t_B-t_A$ between Bob's and Alice's detection times, we construct the coincidence histograms shown in Figure~\ref{fig:coincidence_peaks}. For~each pair of physical detector channels, the~four possible bit combinations correspond to coincidence peaks located at
\begin{equation}\label{eq3}
    D_{ab} - \delta_a \quad \text{for 10,} \qquad D_{ab} \quad \text{for 00,} \qquad
    D_{ab} - \delta_a + \delta_b \quad \text{for 11,} \qquad D_{ab} + \delta_b \quad \text{for 01,}
\end{equation}
where $\delta_a$ and $\delta_b$ are the delays introduced by Alice's and Bob's fibre delay lines, respectively, and~$D_{ab}$ is the time offset between Alice's and Bob's detections that have no delay line. This offset accounts for the fibre, detector and electronic delays of both~channels.

\begin{figure}[H]
    \centering
    \includegraphics[scale=1.1]{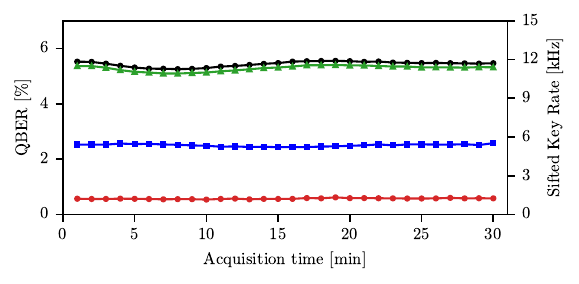}
    \caption{Evolution over time of QBER (blue and red points, left abscissa) and Sifted Key Rate (black and green points, right abscissa). Blue squares: $e_X$; red circles: $e_Z$; black circles: sifted key rate in the $Z$ basis; green triangles: sifted key rate in the $X$ basis. Each point is the result of \SI{60}{\second} of acquisition~time.}
    \label{fig:qber_and_skr}
\end{figure}

Fitting the 16 peak positions of the four channel pairs by weighted least squares, we find delays of \SI{108807.9\pm0.1}{\pico\second} 
 and \SI{108872.6\pm0.3}{\pico\second} for Alice's two arms, $\delta_{a,X}$ and $\delta_{a,Z}$, and~\SI{133388.2\pm0.1}{\pico\second} and \SI{133304.6\pm0.1}{\pico\second} for Bob's, $\delta_{b,X}$ and $\delta_{b,Z}$.
Using the fitted delays, we reconstruct eight logical detection channels from the four physical detector time-tag streams. This reconstruction allows the identification of the logical outcomes associated with the observed coincidence peaks and the classification of experimentally resolved multiple-click events. The~relatively long delays employed in our implementation substantially mitigate detector dead-time effects, allowing the SNSPDs to recover between temporally separated logical outcomes. This enables a closer approximation to the detection statistics of a receiver employing eight independent detectors.
Using the reconstructed dataset and a coincidence window of \SI{150}{\pico\second} centred on each expected peak, we extracted the events used for key sifting. The~resulting error rates were $e_Z = \SI{0.57\pm0.02}{\percent}$ and $e_X = \SI{2.50\pm0.04}{\percent}$, quoted as the mean and standard deviation over 30 acquisition blocks of \SI{60}{\second} each.
Figure~\ref{fig:qber_and_skr} shows the evolution of the QBER and of the sifted key rate over time for~both measurement bases. The~sifted key rate is \SI{11.6\pm0.2}{\kilo\bit/\second} in the $Z$ basis and \SI{11.3\pm0.2}{\kilo\bit/\second} in the $X$ basis, giving a total sifted rate of \SI{22.9\pm0.4}{\kilo\bit/\second}. The~two rates are nearly equal, and they drift together over the acquisition, which points to a common cause such as slow changes in the source's pump laser. A~sifted key rate of this order is typical of current laboratory implementations, but~it is low compared with the pair-emission rate of the EPS of approximately \SI{1}{\mega\hertz}. Several factors significantly reduce the coincidence rate in this apparatus. Only a small slice of the source's emission spectrum is selected by the optical filters. Furthermore, a~number of high-loss components are currently required for operation, notably the high insertion loss tunable optical filter and the beam combiners used for time-multiplexed detection. Given these additional losses, we consider the observed sifted key rate consistent with~expectations.

Table~\ref{tab:bbm92_statistics} summarises the detected-event statistics. Besides~single-click coincidences, the~logical-port reconstruction also reveals multiple-click events, in~which at least one party registers detections corresponding to different logical outcomes within the delay-compensated coincidence windows. Among~the results summarised in Table~\ref{tab:security_proofs_combined}, such events are discarded~\citep{Koashi2008Practical}, assigned random bits~\citep{TsurumaruTamaki2008}, or~excluded by the assumption of binary measurements~\citep{TomamichelLeverrier2017,Lim2021}. Where both same-basis and cross-basis double clicks are treated explicitly~\citep{Gittsovich2014,kawakami2026,nahar2026}, the~analysis is asymptotic or a measurement reduction, and~no composable finite-key bound covers them for a passive BBM92 receiver. However, these events occur in practical implementations and, if~left uncharacterised, could be exploited by an eavesdropper sending multiphoton states. Being able to resolve and count them, as~done here, therefore provides useful input for a complete security~analysis.
\begin{table}[H]

  \caption{Measured coincidence counts over 30 min considering a \SI{150}{\pico\second} coincidence window.}
  \label{tab:bbm92_statistics}
  \sisetup{group-separator = {\,}, group-minimum-digits = 4}
  \begin{tabularx}{\textwidth}{LLR}
    \toprule
    \textbf{Alice} & \textbf{Bob} & {\textbf{Counts}} \\
    \midrule
    Single             & Single             & 82,344,505 \\
    Cross-Basis Double & Single             & 9125 \\
    Single             & Cross-Basis Double & 8289 \\
    Single             & Same-Basis Double  & 3159 \\
    Same-Basis Double  & Single             & 2361 \\
    Cross-Basis Double & Cross-Basis Double & 65 \\
    Cross-Basis Double & Same-Basis Double  & 22 \\
    Same-Basis Double  & Cross-Basis Double & 7 \\
    Same-Basis Double  & Same-Basis Double  & 4 \\
    \bottomrule
  \end{tabularx}
\end{table}
\subsection{From Experimental Data to Security~Guarantees}

The experimental implementation considered here presents two distinct levels of security challenges. The~first concerns imperfections that are already present in a conventional passive BBM92 receiver, including biased basis selection, unequal detection efficiencies and multiple detection events. The~second arises specifically from the time-multiplexed architecture, which introduces detector recovery effects and requires additional timing information to reconstruct measurement outcomes. We examine how these two classes of imperfections relate to the assumptions of existing security proofs and discuss the limitations of their direct application to our experimental~data.

\subsubsection{Passive Basis Selection and Detection-Efficiency~Mismatch}

The results summarised in Table~\ref{tab:security_proofs_combined} address different aspects of the security of practical BBM92 receivers. A~distinction must be made between a security theorem for an abstract quantum measurement and a rigorous reduction of a physical optical receiver to that~measurement. 

For entanglement-based QKD, Koashi~et~al.~\citep{Koashi2008Practical} explicitly analyse multiphoton and multimode optical inputs with threshold detectors but~consider active basis selection and particular detector assumptions. Tsurumaru and Tamaki~\citep{TsurumaruTamaki2008} establish a squashing reduction applicable to BBM92 with actively selected measurements and prescribed random assignment of double clicks. These results demonstrate that multiphoton emission and threshold detection can be treated rigorously, but~they do not directly establish security for the passive four-detector receiver of Figure~\ref{fig:setup}a.

An intermediate result is the passive optical squashing construction of Gittsovich~et~al.~\citep{Gittsovich2014}. It treats a specified balanced passive BB84 receiver, including same-basis multiple clicks and cross-basis clicks, under~particular classical processing rules.~Although~such a reduction may provide an ingredient for a BBM92 security analysis, its applicability cannot be assumed for our receiver, which exhibits an imbalanced basis splitter and unequal transmission through the individual bit~paths.

More recently, Kawakami~et~al.~\citep{kawakami2026} have established asymptotic security directly for passive entanglement-based QKD, including BBM92 with biased basis selection. Their analysis incorporates multiphoton inputs, threshold detectors and cross-basis click statistics. However, the~receiver model assumes equal detection efficiencies for the two bit outcomes within each basis, together with further restrictions on detector parameters. These conditions are not established by our optical characterisation. Furthermore, the~result is asymptotic and therefore does not directly provide a composable finite secret-key length for our experimental~blocks.

Finite-key analyses such as those of Tomamichel and Leverrier~\citep{TomamichelLeverrier2017} and Lim~et~al.~\citep{Lim2021} address a different layer of the problem. They provide composable security bounds for specified entanglement-based protocols, subject to assumptions about their effective measurements, sampling procedures and classical communication. They do not independently derive the required binary measurement model from our complete passive optical apparatus. Consequently, applying such a bound to the measured QBERs requires an additional justification of the receiver measurement and its post-processing.

A particularly important limitation of the entanglement-based model of Ref.~\citep{TomamichelLeverrier2017} is its assumption of deterministic binary measurement outcomes. Assigning random bits to no-detection events would be prohibitive in high-loss optical implementations. Although~their prepare-and-measure analysis allows inconclusive outcomes under an explicit basis-independent detection condition, the~application of an analogous postselection procedure to a practical passive BBM92 receiver requires separate~justification.

The remaining difficulty should not be interpreted as an inability of security theory to accommodate realistic detector imperfections. Recent detector-reduction techniques can treat substantially more general imperfect threshold detectors, including characterised ranges of efficiencies and dark-count parameters~\citep{nahar2026}. Related finite-size techniques also address passive detection with imperfections and memory effects~\citep{Wang2025Passive}, although~their demonstrated applications do not establish the complete BBM92 security claim considered~here. 

A complementary line of research addresses the detection-efficiency mismatch in prepare-and-measure QKD. Trushechkin~\citep{Trushechkin2022} proved asymptotic security of BB84 against general attacks while allowing multiphoton signals at the receiver. Grasselli~et~al.~\citep{Grasselli2025} developed an asymptotic analysis under collective attacks for basis-dependent detection probabilities, using additional measurement statistics enabled by a tunable beam splitter. Tupkary~et~al.~\citep{Tupkary2025Detectors} established a finite-key proof for decoy-state BB84 against coherent attacks with bounded detector-efficiency and dark-count mismatches. Curr\'as-Lorenzo~et~al.~\citep{CurrasLorenzo2026} subsequently introduced a modular finite-key framework for BB84-type protocols that incorporates both source and detector imperfections. Most recently, Ivchenko~et~al.~\citep{ivchenkoSecurityQuantumKey2026} presented a key estimation scheme that allows for arbitrarily large detector-efficiency mismatches in BB84, as~opposed to the bounded approach employed by Tupkary~et~al.~\citep{Tupkary2025Detectors}. These results demonstrate that basis-independent detection is not a universal prerequisite for QKD security. However, none directly establishes security for the passive entanglement-based receiver considered here, with~potentially imperfect measurements at both Alice and Bob. In~particular, Curr\'as-Lorenzo~et~al. identify the extension to entanglement-based protocols with detector mismatches at both users as an open problem within their framework, even though a related extension to passive basis choice at a single receiver has been developed~\citep{Wang2025Passive}. Nor do these results establish security for the timing-derived outcomes and detector recovery effects of our time-multiplexed~receiver.

Finally, attempting to equalise the receiver efficiencies does not automatically resolve the problem. Physical attenuation may simplify the measurement model if the resulting losses can be rigorously represented as trusted, independent pre-detection loss. Randomly discarding a fraction of the registered clicks is not generally equivalent to such attenuation for arbitrary multiphoton inputs. Likewise, a~nominally balanced basis splitter does not remove possible wavelength-, polarisation- or time-dependent detection mismatch. The~appropriate criterion is therefore not equality of the observed count rates alone, but~agreement between the characterised receiver, its click-processing rules and the assumptions of the selected security~proof.

\subsubsection{Timing Reconstruction and Information~Leakage}

As described in Section~\ref{sec:multiplex}, the~present CW implementation does not provide either user with a locally identifiable logical bit outcome for each physical detector click. Instead, the~bit pair is reconstructed from the relative arrival times of Alice's and Bob's detections. This introduces a security concern that is absent from the conventional four-detector configuration, where the bit value is identified directly by the detector that~clicks.

The origin of this concern can be understood from Equation~(\ref{eq3}) and Figure~\ref{fig:coincidence_peaks}a.
For a given pair of physical detector channels, the~four possible bit combinations correspond to four distinct relative detection times. For~example, a~coincidence identified at the offset $D_{ab}$ corresponds to the bit pair $00$, whereas one at $D_{ab}+\delta_b$ corresponds to $01$. Consequently, knowledge of the physical detector channels and sufficiently precise relative timing information allows the corresponding bit pair to be identified. The~experimentally resolved peaks in Figure~\ref{fig:coincidence_peaks}a illustrate this~correspondence.

This observation has a direct cryptographic consequence. If~Alice and Bob publicly disclose timestamps with sufficient resolution to distinguish these coincidence peaks, and~the physical detector-channel labels and calibrated delays are public or otherwise inferable, an~eavesdropper can in principle reproduce the same bit-pair reconstruction. Unlike the usual public announcement of measurement bases in BBM92, such a disclosure would reveal the key-generating outcomes~themselves. 

Importantly, this argument applies to the specified form of public timing disclosure. It does not establish that time multiplexing is intrinsically insecure, nor does it exclude other possible classical processing procedures. It demonstrates, however, that the coincidence-reconstruction method used to analyse our experimental data cannot be directly adopted as a secure BBM92 post-processing procedure without an additional security argument. This issue is distinct from previously demonstrated detector-timing side channels~\citep{LamasLinares2007}, since the timing information in our implementation is used to identify the logical measurement outcomes themselves. Moreover, coincidence selection requires careful analysis when the source is untrusted, since the rules used to pair detections and accept events must not introduce an unaccounted outcome-dependent selection~bias.

Detector recovery introduces an additional, independent complication. Because~two logical bit outcomes share one SNSPD, an~earlier detection can modify the probability of registering a later one and can suppress multiple-click patterns. Delays chosen to allow substantial detector recovery reduce this effect but do not establish a memoryless measurement model. An~applicable security analysis must either account for the history-dependent response or justify a measurement procedure that removes the relevant~dependence.

Overcoming this limitation requires a measurement and classical post-processing procedure that preserves the secrecy of the raw key while providing the information necessary for coincidence identification and parameter estimation. Establishing the security of such a procedure would additionally require a compatible treatment of the effective optical measurement, detector recovery and coincidence selection.
Possible mitigation routes include calibrated spectral filtering and characterisation over the wavelengths accessible to Eve, optical-path balancing or security models that explicitly allow the measured efficiency mismatch, and~timing-announcement rules that do not disclose the logical bits. Longer delays can mitigate detector-recovery effects. None of these measures alone establishes security without an analysis of the complete optical measurement, public transcript and coincidence-selection~procedure.

Accordingly, the~present time-multiplexed measurements demonstrate the expected entanglement-based correlations, but~no composable finite-key security claim follows from the existing proofs considered here. Table~\ref{tab:security_proofs_combined} compares the relevance of these results to both receiver~architectures.

\section{Discussion and~Conclusions}

Our experimental investigation highlights the difficulty of translating BBM92 measurements into a provably secure cryptographic key. While the observed QBERs are low, these results cannot be used to establish a key with composable security. For~the conventional passive receiver of Figure~\ref{fig:setup}a, existing security analyses address several relevant imperfections, including multiphoton inputs, threshold detection, passive basis selection and detector-efficiency mismatch~\citep{Koashi2008Practical,Gittsovich2014,kawakami2026,nahar2026}. However, a~rigorous connection between the complete physical receiver, its classical post-processing and an applicable finite-key BBM92 theorem has not yet been~demonstrated.

The importance of careful experimental characterisation is exemplified by the landmark Micius experiment by Yin~et~al.~\citep{Yin2020Micius}, which demonstrated entanglement-based QKD over 1120~km. This remarkable achievement involved extensive receiver characterisation, spectral and temporal filtering, and~countermeasures against detector side-channel attacks, setting an important benchmark for practical quantum communications.
In parallel, theoretical developments, including the ones reported in Table~\ref{tab:security_proofs_combined}, have substantially strengthened the security foundations of entanglement-based QKD. These analyses establish rigorous security guarantees under well-defined assumptions about the quantum measurements and classical post-processing. Connecting these theoretical models to the complete physical operation of a practical receiver remains an important task, particularly as increasingly sophisticated experimental architectures introduce additional imperfections and processing~requirements.

Our time-multiplexed receiver, Figure~\ref{fig:setup}b, introduces an additional difficulty: with a CW-pumped source, the~logical
bit values are reconstructed from the relative detection times of Alice and Bob, potentially revealing information through public timing announcements. Furthermore, detector recovery introduces history-dependent detection efficiencies. A~possible remedy is to provide each user with an independent timing reference, for~example, through a pulsed source, enabling local identification of the bit values. However, the~resulting receiver would still require a security analysis accounting for detector recovery, multiple clicks and coincidence~selection.

While our findings do not imply that passive BBM92 or time multiplexing is intrinsically insecure, they highlight the need to establish a more complete correspondence between the physical implementation and the assumptions of a security proof. This may be achieved either by engineering receivers to satisfy existing security models or by extending current proofs to accommodate realistic optical imperfections. Building on the present findings, our future work will focus on improving the temporal analysis and post-processing of entanglement-based QKD measurements, with~the aim of retaining the practical advantages of time multiplexing while establishing a more rigorous foundation for secure key~extraction.

\vspace{6pt}

\authorcontributions{Conceptualisation, K.P.S. and M.L.; methodology, K.P.S., E.F., P.P. and M.L.; software, K.P.S. and E.F.; validation, K.P.S. and E.F.; formal analysis, E.F., P.P. and~M.L.; investigation, K.P.S. and E.F.; data curation, E.F.; writing---original draft preparation, K.P.S. and E.F.; writing---review and editing, K.P.S., E.F., P.P. and M.L.; visualisation, K.P.S. and E.F.; supervision, P.P. and M.L.; project administration and funding acquisition, M.L. All authors have read and agreed to the published version of the manuscript.}


\funding{This 
 research was partially supported by the Engineering and Physical Sciences Research Council (EPSRC) through the UK Quantum Communications Hub (Grant No.~EP/T001011/1) and the UK Integrated Quantum Networks Hub (Grant No.~EP/Z533208/1), as well as~by an EPSRC Impact Acceleration Account (IAA) Small Award administered by the University of York (Grant No.~EP/X525856/1).
E.F. acknowledges support from an Erasmus+ Traineeship mobility grant co-funded by the European Union (project No.~2024-1-IT02-KA131-HED-000225621, Erasmus+ National Agency), with~additional funding from the Italian Ministry of University and Research (MUR, Fondo 
 per il sostegno dei~giovani) and the University of Camerino, and~from a mobility grant of the School of Advanced Studies ``Carlo Urbani'', University of Camerino. P.P. acknowledges support from the Entanglement Enhanced Quantum Integrated Networks (EQUIN) project, funded by Innovate UK and the Engineering and Physical Sciences Research Council (EPSRC) under Grant No.~10150561, during~the initial stages of this~research.}

\dataavailability{The raw data supporting the conclusions of this article will be made available by the authors on request.}

\acknowledgments{We thank Giovanni 
 Di Giuseppe and Rupesh Kumar for the valuable support and guidance provided to E.F. throughout this~work.}

\conflictsofinterest{The authors declare no conflicts of~interest.} 

\isPreprints{}{
} 

\reftitle{References}

\isPreprints{}{
} 
\end{document}